\documentclass[aps,prb,preprint,superscriptaddress]{revtex4}
\usepackage{bm}
\usepackage{ulem}
\usepackage[dvipdfmx]{graphicx}
\usepackage{subfig}
\usepackage{mediabb}
\usepackage{amsmath}
\usepackage{amssymb}
\usepackage{xcolor}
\usepackage{url}
\usepackage{mathtools}

\def\*#1{\mathbf{#1}}

\newcommand{\eq}[1]{(\ref{#1})}

\newcommand{\lw}[1]{\smash{\lower2.ex\hbox{#1}}}

\newcommand{\RR}{\mathbb{R}}

\newcommand{\EE}{\mathbb{E}}
\newcommand{\VV}{\mathbb{V}}

\newcommand{\cX}{{\cal X}}

\begin{document}

\title{Active-learning-based efficient prediction of {\it ab initio} atomic energy: 
a case study on a Fe random grain boundary model with millions of atoms}

\author{Tomoyuki Tamura}
\email{tamura.tomoyuki@nitech.ac.jp}
\affiliation{Department of Physical Science and Engineering, Nagoya Institute of Technology, Nagoya,  466-8555, Japan}
\affiliation{Center for Materials research by Information Integration, National Institute for Materials Science (NIMS), Tsukuba 305-0047, Japan}

\author{Masayuki Karasuyama}
\email{karasuyama@nitech.ac.jp}
\affiliation{Department of Computer Science, Nagoya Institute of Technology, Nagoya, 466-8555, Japan}
\affiliation{Center for Materials research by Information Integration, National Institute for Materials Science (NIMS), Tsukuba 305-0047, Japan}
\affiliation{PRESTO, Japan Science and Technological Agency, 4-1-8 Honcho, Kawaguchi, Saitama 332-0012, Japan}

%\maketitle

\begin{abstract}
We have developed a method that can analyze large random grain boundary (GB) models 
	with the accuracy of density functional theory (DFT) calculations using active learning.
It is assumed that the atomic energy is represented 
	by the linear regression of the atomic structural descriptor.
The atomic energy is obtained through DFT calculations using a small cell extracted from a huge GB model, 
	called replica DFT atomic energy.
The uncertainty reduction (UR) approach in active learning is used 
	to efficiently collect the training data for the atomic energy.
In this approach, atomic energy is not required to search for candidate points; 
	therefore, sequential DFT calculations are not required.
This approach is suitable for massively parallel computers that can execute a large number of jobs simultaneously.
In this study, we demonstrate the prediction of the atomic energy 
	of a Fe random GB model containing one million atoms using the UR approach
	and show that the prediction error decreases more rapidly compared with random sampling. 
We conclude that the UR approach with replica DFT atomic energy is useful for modeling huge GBs
	and will be essential for modeling other structural defects. 
			
\end{abstract}

\maketitle

\clearpage

% --------------------------------------------------
\section{Introduction} \label{sec:introduction}

A grain boundary (GB) is the interface between two grains or crystals in a polycrystalline material. 
Atomic configurations and chemical bonds near GBs are distinct from those of the bulk crystal. 
Thus, the electrical properties of materials with GBs can greatly differ from those of a single crystal, 
	and GBs govern a wide range of material properties \cite{Sutton1995}.
A majority of the GB research is based on the coincidence site lattice (CSL) theory \cite{Kronberg1949};
	a CSL GB is a simplified model with regularity and is usually characterized by the $\Sigma$ value, 
	which is defined as the reciprocal of the density of the coincidence sites. 
To understand the atomic and electronic structures of CSL GBs, 
	experiments using high-resolution electron microscopies 
	and computer simulations using empirical potentials and first-principles calculations have been utilized.
However, most GBs in actual materials are random, with no regularity, and have local amorphous structures.
As it is difficult to identify atomic arrangements in amorphous phases using experimental observations, 
	computer simulations play a major role.
To study random GBs based on simulations, 
	a large supercell containing a large number of atoms is required, which results in a high computational cost.
Although a classical molecular dynamics (MD) simulation using a model that contains billions of atoms can be performed 
	 using the latest supercomputers,
	results obtained from the simulation depend on the parameters associated with the empirical potentials.
Therefore, first-principles calculations are required, 
	but modeling random GBs is not realistic, as it incurs huge computational costs,
	which include computational time and memory size.
First-principles plane-wave density functional theory (DFT) calculations are widely used to 
	identify defects in materials.
Various $O(N)$ DFT methods have been developed \cite{siesta, openmx, conquest}, 
	but it remains impossible to model a huge GB.

Iron and its alloys, because of their high strength and toughness, play important roles as structural materials
	in industries, infrastructures, and our daily lives.
The high strength and toughness of these polycrystalline materials are strongly affected by their GBs 
	\cite{Sutton1995}.
Therefore, it is necessary to clarify the correlation 
	between GB microstructures and mechanical properties.
The GBs in iron and its alloys have been studied extensively for a long time, 
	being one of the most studied GBs to date.
Recently, Shibuta {\it et al.} performed classical MD simulations 
	using a model containing one million and one billion atoms to investigate the nucleation process of pure Fe
	\cite{Shibuta2016,Shibuta2017}.
As a result, huge random GB models have been obtained.

In plane-wave DFT calculations, the total energy is obtained as the average value of the whole system.
Conversely, an attempt has been made to analyze the local physical properties 
	using the local energy obtained by dividing the total energy into local regions \cite{Shiihara2010}.
Using this local-energy analysis, a large amount of local energy can be extracted from the DFT calculation of one system.
However, the computational cost of the DFT calculation for a GB model is high; thus, 
	exhaustive investigation is not realistic for large $\Sigma $ CSL GBs.
We have developed an efficient scheme to predict the GB energy, 
	where the correlation between the local environment and the DFT atomic energy for GB atoms is 
	estimated from a few small $\Sigma $ CSL GBs 
	with a linear regression model in machine learning,
	and the atomic energy and its sum, namely the GB energy,  
	are predicted using the local environment information as a descriptor and learned parameters
	for any CSL GB \cite{Tamura2017}.
We applied this scheme to the fcc-Al [110] tilt CSL GB and obtained good prediction results.
However, as the DFT local-energy analysis requires a supercell calculation under periodic conditions, 
	it cannot be directly applied to random GB models.

\begin{figure}[t]
\begin{center}
\includegraphics[scale=0.4]{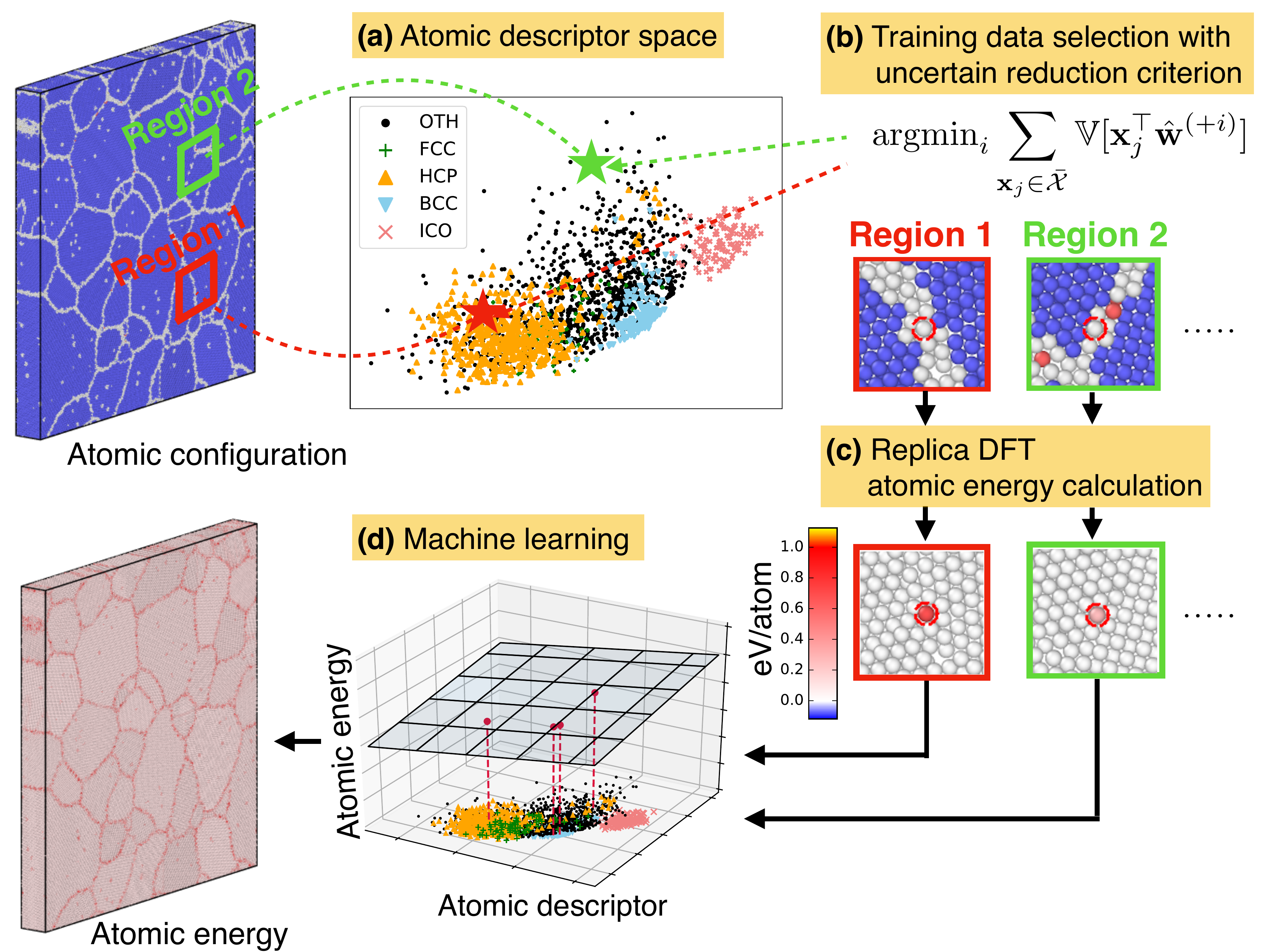}
\caption{
	Schematic of the non-sequential prediction of the {\it ab initio} atomic energy of a random GB model.
	(a) Atomic descriptors for all atoms are calculated. 
	(b) Training data are selected based on the uncertain reduction criterion.
	(c) Replica DFT atomic energy calculations are performed non-sequentially for selected atoms.
	(d) Using the atomic descriptors and the calculated atomic energy, the machine-learning model parameters are optimized.
	Then, the DFT-based atomic energy of a random GB model can be predicted quickly without time-consuming computations.
 }  
\label{fig:flowchart}
\end{center}
\end{figure}

In this study, we have developed a machine-learning-based method to predict the atomic energy 
	of a huge random GB model using the DFT local-energy analysis.
A tractable replica supercell that includes the surrounding atomic arrangements is constructed 
	for each atom in a random GB model,
		and the training data of the DFT atomic energy are collected. 
Similar to our previous scheme \cite{Tamura2017},
	the atomic energy is predicted using local environment information and learned parameters.
Although this strategy is effective, the selection of the training data set can have a significant effect on the prediction accuracy.
Therefore, we use an {\it active-learning} (AL) approach \cite{Settles:2012}, which has been widely studied 
	in the machine-learning community, to select an appropriate training data set.
In particular, we construct the training data set so that the uncertainty of the prediction 
	over the entire huge GB model can be minimized.
The application of the AL approach to the on-the-fly learning of interatomic potentials fitted to DFT results,
	such as the neural network potentials (NNP) \cite{Behler2007,Behler2011,Behler2014},
	the Gaussian approximation potentials (GAP) \cite{Bartok2010,Szlachta2014,Bartok2013, Bartok:2015iw,De:2016ia, 
	Deringer2017},
	and the moment tensor potentials (MTP) \cite{Shapeev2016},
	has been proposed for long time-scale MD simulations of systems containing hundreds of atoms 
	\cite{Frederiksen2004, Csanyi2004, Mueller2012, Artrith2012, Behler2014, Li2015, 
	Botu2015, Bartok2010,Podryabinkin2017, Peterson2017, Bartok2018, Imbalzano2018, Podryabinkin2019, 
	Bernstein2019, Jinnouchi2019a, Jinnouchi2019b, Zhang2019, Vandermause2020, Mueller2020}.
As described above, it is impossible to perform DFT calculations for large length-scale systems, 
	such as the present random GB model, 
	and the existing on-the-fly AL approaches cannot be applied directly to those systems.		
In Section~\ref{sec:discussions}, we describe further details of this relationship with on-the-fly AL approaches.
Figure~\ref{fig:flowchart} shows a schematic of our proposed procedure, in which a machine-learning model is built 
	based on an atomic descriptor space.
Based on the huge Fe GB model, we show that our strategy can rapidly decrease the prediction error
	compared with simple random sampling approaches. 
An important point is that the DFT calculation is not necessary to select candidate points 
	because the uncertainty criterion does not depend on the atomic energy.
In other words, sequential DFT calculations are not necessary, and it is possible to execute many calculations
	simultaneously using a massively parallel computer.

\section{Method} \label{sec:method}

\subsection{GB model} 

\begin{figure}[t]
\begin{center}
\includegraphics[scale=0.75]{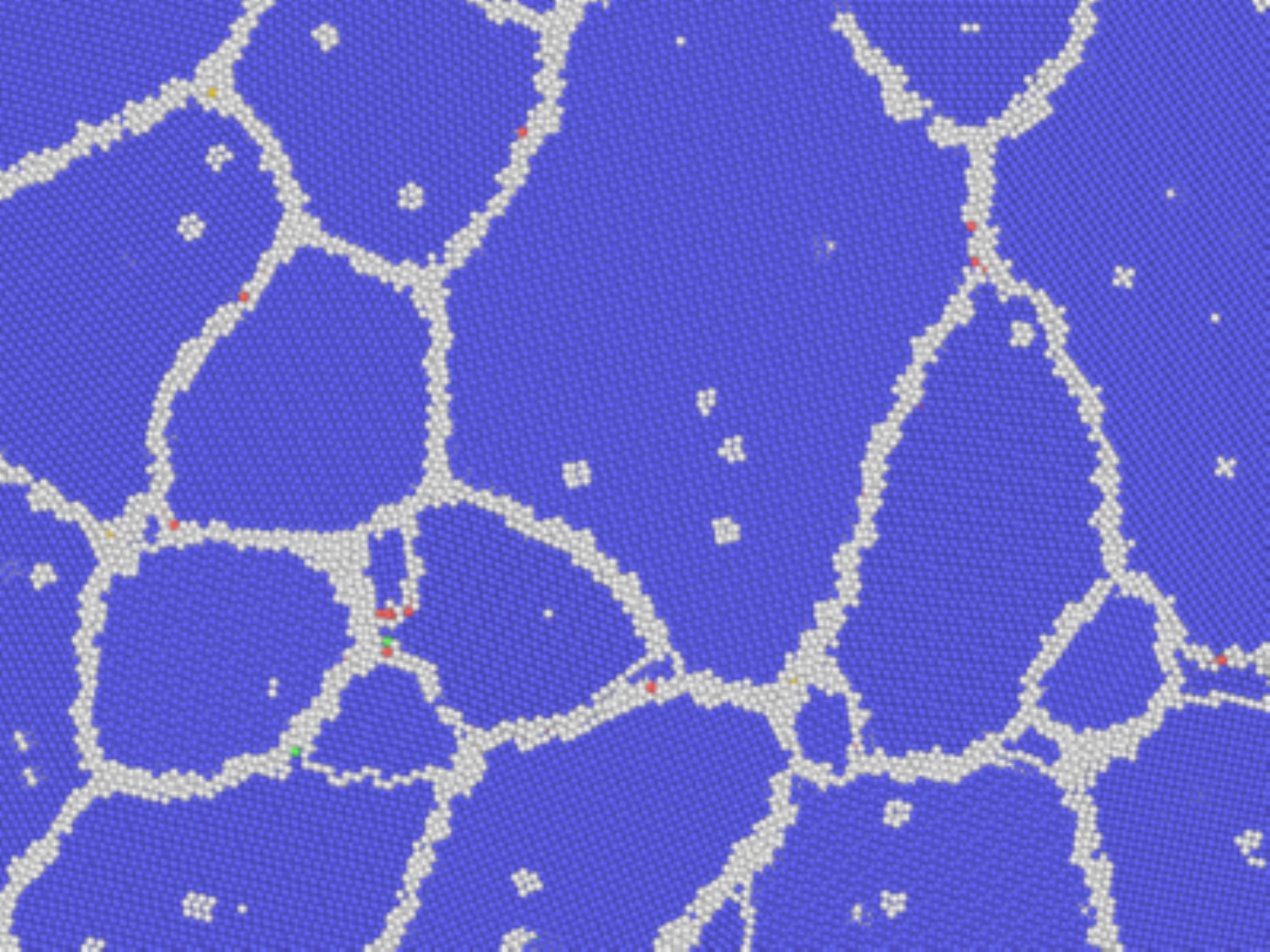}
\caption{Atomic configurations on a plane.  
	Blue, yellow, red, green, and gray represent atoms with BCC, ICO, HCP, FCC, 
		and OTH configurations, respectively, 
		as defined by the adaptive CNA.}
\label{fig:CNA}
\end{center}
\end{figure}

In this paper, we use a model containing 1,037,880 Fe atoms at 1,400 K obtained in Ref. \onlinecite{Shibuta2016}.
The atomic configuration is relaxed with the embedded atom method (EAM) \cite{Song2013}.
The obtained atomic configuration is visualized using the Open Visualization Tool (OVITO) \cite{OVITO},
	as shown in Fig.~\ref{fig:CNA}. 
A common neighbor analysis (CNA) is then performed to identify atomic configurations. 
The adaptive CNA \cite{Stukowski2012}, which employs variable cutoff distances, 
	distinguishes atomic configurations precisely as face-centered cubic (FCC), 
	hexagonal closed pack (HCP), body-centered cubic (BCC), icosahedron (ICO), 
	and unknown (OTH) coordination structures.
Most atoms are BCC and OTH at GBs.
Only a few atoms are ICO, HCP, or FCC.

\subsection{Calculation of DFT-based atomic energy} 

In plane-wave DFT calculations, the total energy is obtained as the average value of the whole system.
The supercell is divided into Bader regions around the atom \cite{Bader1990}, 
	and the integral value in that region denotes the atomic energy $E_{i}^{\mathrm{atom}}$.
	\begin{equation}
		E^{\mathrm{tot}}=\sum_{i} \int_{V_{i}^{\mathrm{Bader}}} \epsilon (\mathbf{r}) \, d\mathbf{r}
		=\sum_{i}E_{i}^{\mathrm{atom}} . 
	\end{equation}
This local-energy analysis scheme, incorporated in the computational code QMAS \cite{qmas}
	based on the projector augmented-wave (PAW) method \cite{Blochl1994,Holzwarth1997,Kresse1999}, 
	has already been applied to defect systems such as fcc-Al (111) surfaces \cite{Shiihara2010}, 
	fcc-Al and fcc-Cu [110] GBs \cite{Wang2013, Wang2015, Wang2017}, 
	and bcc-Fe [110] GBs \cite{Somesh2014, Somesh2014b}.	
We can obtain the unique local energies since the gauge-dependent terms are integrated to be zero 
	and the other ambiguities are averaged out, as discussed in Ref. \onlinecite{Shiihara2010}.
We use the spin-polarized generalized gradient approximation \cite{PBE} for the exchange--correlation functional
	and a cut-off energy of 544 eV for the valence wave function.

The cubic cell centered on the target atom is extracted from the huge GB model.
We fixed the cell size to $10 \times 10 \times 10 \  \mathrm{\AA^ {3}}$.
As a very close atomic pair occurs near the edge of the cell,
	those with interatomic distances of $<$2.2 \AA\ have been removed.
The atomic energy is calculated using the DFT by fixing the atomic configuration,
	and the bulk energy is subtracted per atom.
As metallic bonding has a large screening effect, 
	the atomic energy of the target atom can be obtained with accuracy using a small cell.
We call this atomic energy the {\it replica DFT atomic energy}.
Certainly, we can use a cluster model to obtain the atomic energy.
In general, the convergence of the self-consistent calculation for the system that includes the surface is slow. 
We have checked for an atom in the bulk-like region
	that neither an increase of the box size (from 9 to 14 \AA) 
	nor an increase of the minimum of the interatomic distance
	(from 1.9 to 2.4 \AA)
	changes the atomic energy by more than 1 mRy, 
	which was used for a tolerance of the pseudopotentials \cite{Rappe1990}.

\subsection{Regression model for atomic energy of the GB model} 

We assume that the atomic energy of the $i$-th atom 
	$E_{i}^{\mathrm{atom}}$
	can be represented as
	\begin{equation}
		E_{i}^{\mathrm{atom}} = \mathbf{x}^{\top}_{i}\mathbf{w}^{*},
	\end{equation}
	where 
	$\mathbf{x}_{i} \in \RR^{d}$
	is the $d$-dimensional structural descriptor vector and 
	$\mathbf{w}^{*} \in \RR^{d}$
	is the unknown parameter vector.
The actual observation of the atomic energy based on the DFT calculations 
	$y^{\mathrm{DFT}}_{i}$ is assumed to contain independent Gaussian noise
	\begin{equation}
		y^{\mathrm{DFT}}_{i} = \mathbf{x}^{\top}_{i}\mathbf{w}^{*}+\epsilon,
	\end{equation}
	where $\epsilon \sim N(0,\ \sigma^{2})$ and $\sigma^2$ is the variance.
Suppose that 
	$\mathbf{X} \in \RR^{n \times d}$ and $\mathbf{y} \in \RR^{n}$
	are the training data set consisting of $n$ instances.
The $i$-th row of $\mathbf{X}$ is the descriptor vector $\mathbf{x}^{\top}_i$, and the $i$-th element of $\mathbf{y}$ 
	is the calculated atomic energy $y^{\mathrm{DFT}}_{i}$.
Let $\hat{\mathbf{w}}$ be the parameter vector estimated by ridge regression.
Ridge regression minimizes the following objective function with a regularization parameter $\lambda$:
	\begin{equation}
		L= \| \mathbf{y} -\mathbf{X} \mathbf{w} \|^{2}_{2}+\lambda \| \mathbf{w} \|^{2}_{2},
	\end{equation}
	for which the minimizer is written as
	\begin{equation}
		\hat{\mathbf{w}}= \mathbf{M}^{-1}\mathbf{X}^{\top}\mathbf{y},
	\end{equation}
	where $\mathbf{M}\equiv\mathbf{X}^{\top} \mathbf{X} + \lambda \mathbf{I}$ 
	with the identity matrix $\*I \in \RR^{d \times d}$.
Using the estimated $\hat{\mathbf{w}}$, a prediction for the $j$-th atomic energy in the GB model
	can be obtained as
	\begin{equation}
		E_{j}^{\mathrm{atom}} \approx \mathbf{x}^{\top}_{j}\hat{\mathbf{w}}.
	\end{equation}
For the structural descriptor $\mathbf{x}$ of each atom, we employed the smooth overlap of atomic positions 
	(SOAP) \cite{Bartok:2015iw,De:2016ia}.
The SOAP was calculated using the GAP suit code \cite{Bartok2010, Bartok2013} 
	with the number of radial functions $n_{\mathrm{max}}$=10, 
	the angular momenta $l_{\mathrm{max}}$=6, 
	and the cutoff distance $r_{\mathrm{cut}}$=4.0 \AA.
Since the computation of the SOAP is easier than DFT calculations, 
	the SOAP for all atoms in the GB model can be computed.
In our previous study \cite{Tamura2017}, 
	we verified that the SOAP can accurately predict the atomic energy of the fcc-Al GB, 
	in which there occurred a remarkable charge redistribution and a bond reconstruction between interface atoms 
	with reduced coordination numbers.

\subsection{Sampling training data with active learning} 

To estimate $\hat{\mathbf{w}}$, 
	we assume that the $n$ atomic energy values $\mathbf{y}$ are already calculated as the training data.
As the computational cost of DFT calculations is expensive, 
	the possible numbers of $n$ are usually much smaller than the number of atoms in the GB model.
The prediction accuracy of the resulting model depends on the selection of the $n$ training points.
{\it Active learning} \cite{Settles:2012} is a framework that provides sampling schemes of training data 
	for machine-learning algorithms.
Here, we introduce an active learning strategy that reduces the uncertainty of prediction for the GB model.

Let $\cX$ be the set of $n$ training inputs ${\mathbf{x}}_{i}$, and $\bar{\cX}$ be the set of all ${\mathbf{x}}_{i}$
	in the entire GB model.
The prediction for the $j$-th atom $\*x_j \in \bar{\cX}$ is given as
	\begin{equation}
		\hat{y}_j
	  	= \*x_j^\top \hat{\mathbf{w}}
	  	= \*x_j^\top \left (\mathbf{M}^{-1}\mathbf{X}^{\top}\mathbf{y} \right ).
	\end{equation}
To determine effective training samples, we evaluate the uncertainty of the current regression prediction.
Let $\VV[\*a] = \EE[ (\*a - \EE[\*a]) (\*a - \EE[\*a])^\top]$ be the variance--covariance matrix of a random vector 
	$\*a \in \RR^n$, where $\EE$ is the expectation.
 Using $\VV[\*c^\top \*a] = \*c^\top \VV[\*a] \*c$
	for a constant vector $\*c \in \RR^n$, the variance of the prediction for the $j$-th atom is
	\begin{align*}
		\VV[\hat{y}_j] 
	 	& = 
	 	\VV[ \*x_j^\top \mathbf{M}^{-1}\mathbf{X}^{\top}\mathbf{y}  ]
	 	\\
		& = 
	 	\*x_j^\top \mathbf{M}^{-1}\mathbf{X}^{\top}
	 	\VV[ \mathbf{y}  ]
		\mathbf{X} \mathbf{M}^{-1} \*x_j.
	\end{align*}
As the noise term $\epsilon$ is assumed to be independent for each $i$ in 
	$y_i^{\rm DFT} = \*x_i^\top \*w^{*} + \epsilon$, we see $\VV[ \mathbf{y}] = \sigma^2 \*I$.
Then, we obtain 
	\begin{equation}
		\mathbb{V}[\hat{y}_j]
	  	= \sigma^{2} \*x_j^\top \left (\mathbf{M}^{-1}\mathbf{X}^{\top}\mathbf{X} \mathbf{M}^{\mathrm{-1}}\right ) \*x_j.
	 	\label{eq:var}
	\end{equation}
The right-hand side does not contain $\*y$, 
	which means that the prediction variance does not depend on the calculated atomic energy.
Note that the expectation $\EE$ is taken over $\epsilon$ because other variables are regarded as constant in the standard linear regression modeling \cite{Hastie:2009}.
		
Suppose that we add a new candidate ${\mathbf{x}}_{i} \in \bar{\cX} - \cX$ into the training data,
	and $\hat{\mathbf{w}}^{(+i)}$ is the regression coefficient vector 
	``after'' adding ${\mathbf{x}}_{i}$ into the training data.
As $\*M$ and $\*X$ are changed by the addition of $\*x_i$, the variance of prediction  
	with the updated coefficient vector is given as
	\begin{equation}
		\mathbb{V}[\*x^\top_j \hat{\mathbf{w}}^{(+i)}]
		= \sigma^{2} \*x_j^\top
		\left (\mathbf{M}+\mathbf{x}_{i}\mathbf{x}_{i}^{\top}\right )^{-1}
		\left (\mathbf{X}^{\top}\mathbf{X}
		+\mathbf{x}_{i}\mathbf{x}_{i}^{\top} \right )
		\left (\mathbf{M}+\mathbf{x}_{i}\mathbf{x}_{i}^{\top}\right )^{-1} \*x_j.
		\label{eq:var+}
	\end{equation}				
Then, by summing the updated variance values of all the atoms, 
	we obtain the total uncertainty in the prediction analysis for the GB model 
	after the addition of $\*x_i$ into the training data.
	\begin{equation}
		\sum_{ \mathbf{x}_{j} \in \bar{\cX}} 
		\mathbb{V}[ \mathbf{x}_{j}^{\top} \hat{\mathbf{w}}^{(+i)} ].
		\label{eq:acq}
	\end{equation}
We iteratively add $i$ which minimizes this score 
	to the training data, so that the resulting regression model 
	has smaller prediction uncertainty 
	for the entire GB model 
(Note that for this purpose, $\sigma$ is not necessary because it is common for all $i$).
This method is called {\it uncertainty reduction} (UR).
An important property of UR is that it does not require $\mathbf{y}$ 
	because the variance of Eq.~(\ref{eq:var+}) does not depend on $\*y$.
 We can determine a set of candidates before performing DFT calculations.
Therefore, DFT calculations for the training data set can be performed in parallel.
		
\section{Results} \label{sec:results}

\subsection{CNA analysis with principle component analysis} 

\begin{figure}[t]
\begin{center}
\includegraphics[scale=0.5]{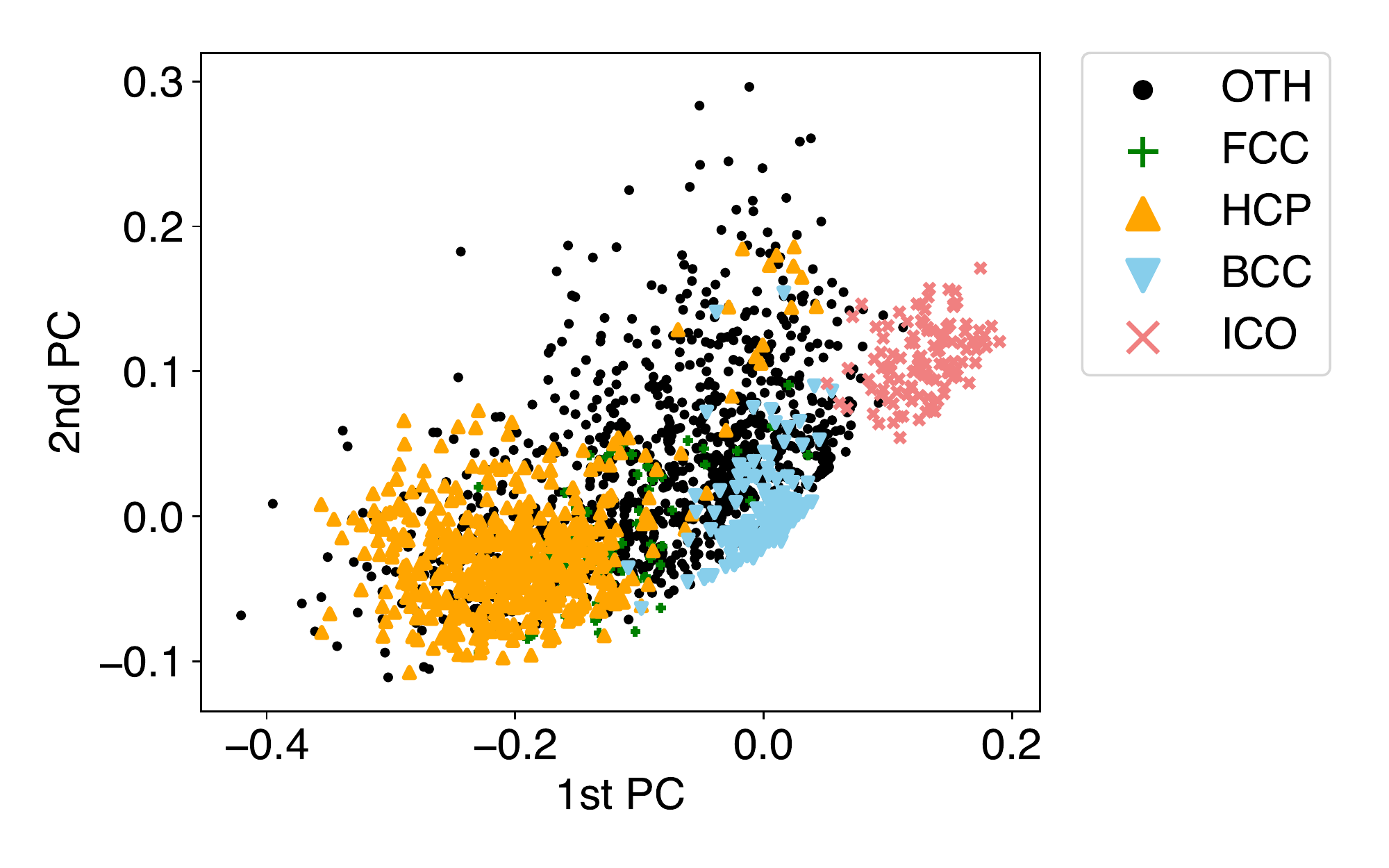}
\caption{PCA analysis of the SOAP descriptor, projected onto the plane of the first two PCs. 
	Data points are the color-coded labels of the CNA structure.}
\label{fig:PCA}
\end{center}
\end{figure}

We generated a $386$-dimensional SOAP vector for each atom in the GB model.
We applied {\it principal component analysis} (PCA) to the original SOAP vector, by which the dimensions were 
	reduced to $39$, keeping $99.99$\% of the original variance.
Figure~\ref{fig:PCA} shows the first two principal components (PCs) with the CNA structure.
This two-dimensional plot contains 92\% of the variance of the original $386$-dimensional space 
	(the first PC contains 68\% of the variance, and the second PC contains 24\% of the variance).
For each structure type, we plot at most $1000$ points randomly chosen from the GB model 
	(if a specific type has less than $1000$ points, all the points in that type are plotted).
We can see that BCC, ICO, FCC, and HCP are concentrated around different centers
	(FCC and HCP are distributed around a similar location because of their structural similarity).
OTH spreads out entirely and partially overlaps with the other known types, 
	although it is also distributed at locations where no known structures exist. 

\subsection{Active learning results} 

\subsubsection{Training and test data settings} 

We evaluated three sampling strategies to create a training data set.
The first set contains $150$ training instances, selected by UR.
The second set, called Random~1, contains $150$ instances randomly selected from all the atoms except for BCC 
	(which we call Non-BCC).
The third set, called Random~2, contains $10$ randomly selected instances from BCC and $140$ instances from Non-BCC.
To create the test data set for performance evaluation regarding unseen atoms, we first define the coordination number 
	as the number of atoms less than $r_\mathrm{cut}$.
For the BCC lattice, the 8 first- and 6 second-nearest neighbors must be considered, and 
	a local cutoff is set halfway between the second and third BCC coordination shells.
	\begin{equation}
		r_{\mathrm{cut}}=\frac{1+\sqrt{2}}{2} a_{\mathrm{bcc}},
		\label{eqn:rcut}
	\end{equation}
 	where local $a_{\mathrm{bcc}}$ is computed using the 14 nearest neighbors as 
	\begin{equation}
		a_{\mathrm{bcc}} ^{\mathrm{local}} 
		=\frac{1}{2} \left [ \frac{2}{\sqrt{3}} \frac{\sum_{j=1}^{8}|\mathbf{r}_{j}|}{8} 
		+ \frac{\sum_{j=9}^{14}|\mathbf{r}_{j}|}{6} \right ].
	\end{equation}		
We randomly selected $10$ atoms from different coordination numbers 11$\sim$18, 
	which resulted in $80$ atoms in total.
We define the set of the test atoms as $\cX_{\rm Test}$.

\subsubsection{Comparison of atomic energy with DFT} 
		
\begin{figure}[t]
\begin{center}
\subfloat[Random 1]{     
	\includegraphics[scale=0.4]{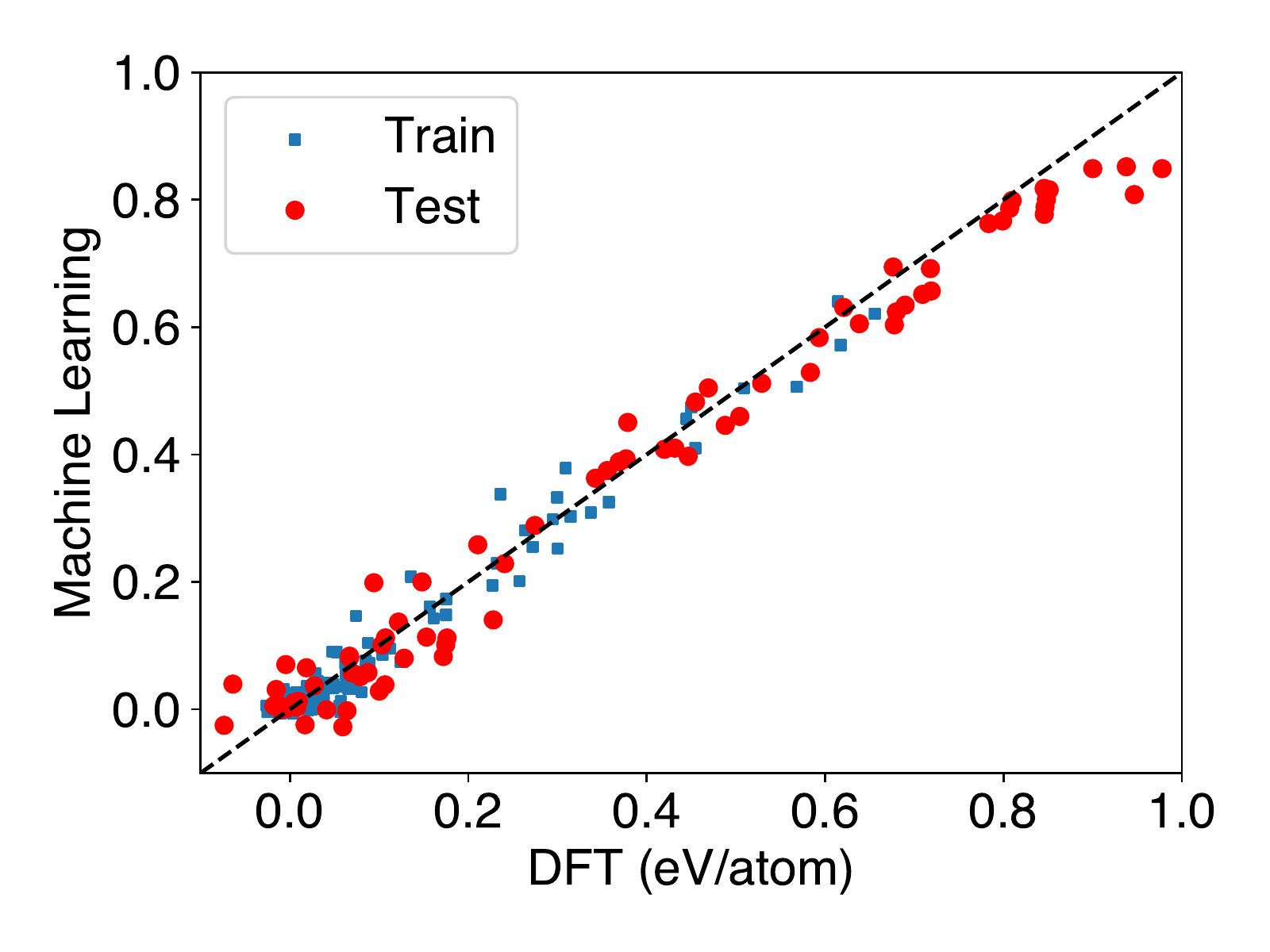}
     }
\subfloat[UR]{     
	\includegraphics[scale=0.4]{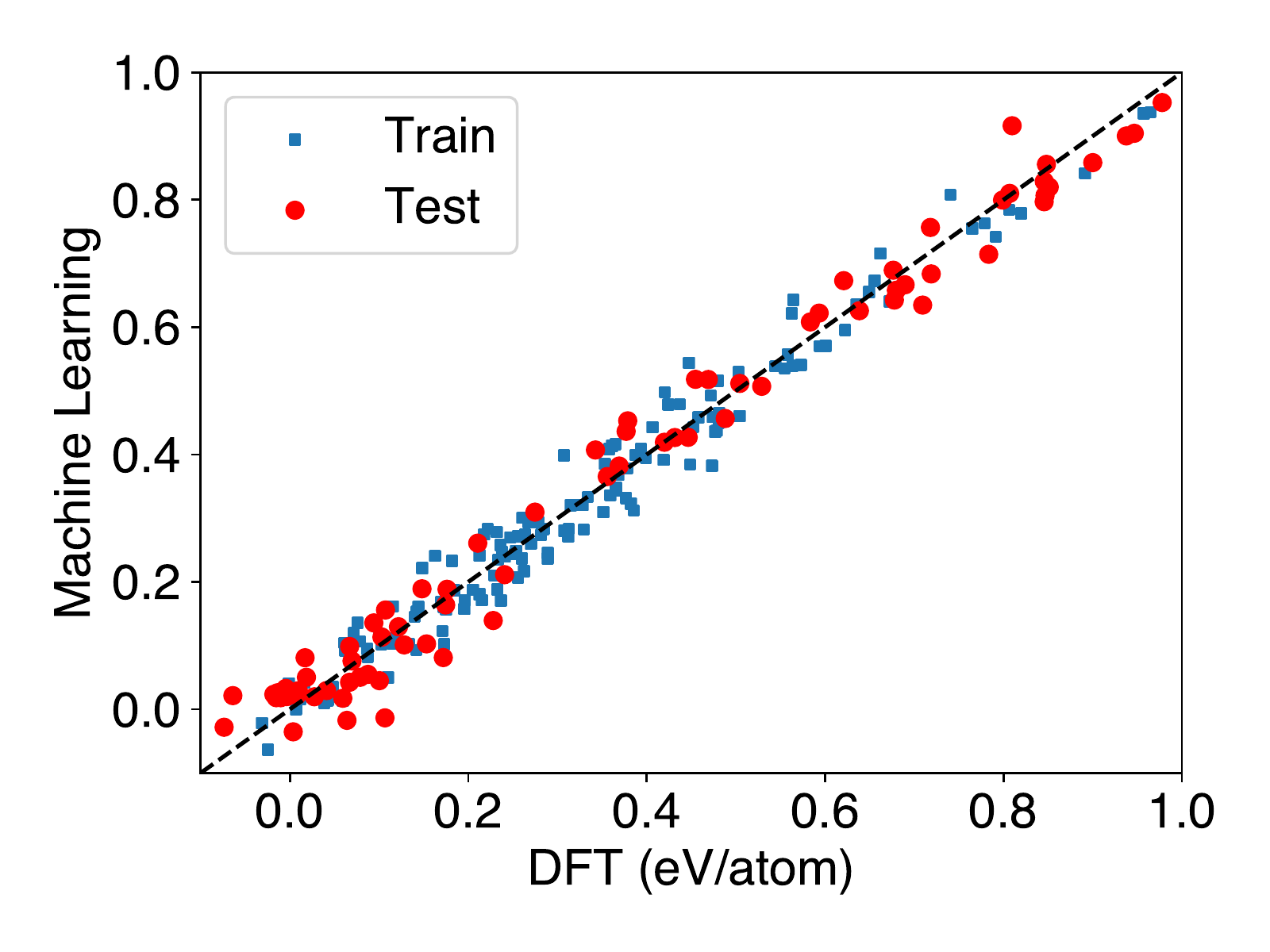}
     }
\caption{
	Comparison between the DFT and predicted values of the atomic energy (eV/atom).
}
\label{fig:Prediction}
\end{center}
\end{figure}

Figure~\ref{fig:Prediction} shows the prediction of Random~1 and UR.
We measure the root mean squared error (RMSE): 
	\begin{equation}
		\sqrt{\sum_{\mathbf{x}_{i} \in \cX_{\rm Test}} (y^{\mathrm{DFT}}_{i} - \mathbf{x}^{\top}_{i} 
		\hat{\mathbf{w}})^2 / |\cX_{\rm Test}| },
	\label{eqn:RMSE}
	\end{equation}
	and the maximum absolute error (MAE):
	\begin{equation}
		\max_{\mathbf{x}_{i} \in \cX_{\rm Test}} |y^{\mathrm{DFT}}_{i} - \mathbf{x}^{\top}_{i} \hat{\mathbf{w}}|.
	\end{equation}
The RMSE was 0.055~eV/atom for Random~1 and 0.044~eV/atom for UR.
The MAE was 0.212~eV/atom for Random~1 and 0.120~eV/atom for UR.
First, both predictions, Fig. 4(a) and (b), were surprisingly accurate because the training data set had only $150$ atoms, 
	which is $<$0.02\% ($0.00014 \approx 150/1037880$) of the entire GB model. 
Further, UR outperformed Random~1 related to both RMSE and MAE.
In particular, Random~1 does not have training instances for larger-energy regions ($>0.7$~eV/atom).
This lack of training data negatively affected the prediction accuracy of Random~1 for larger atoms.

\begin{figure}[t]
\begin{center}
\includegraphics[scale=0.4]{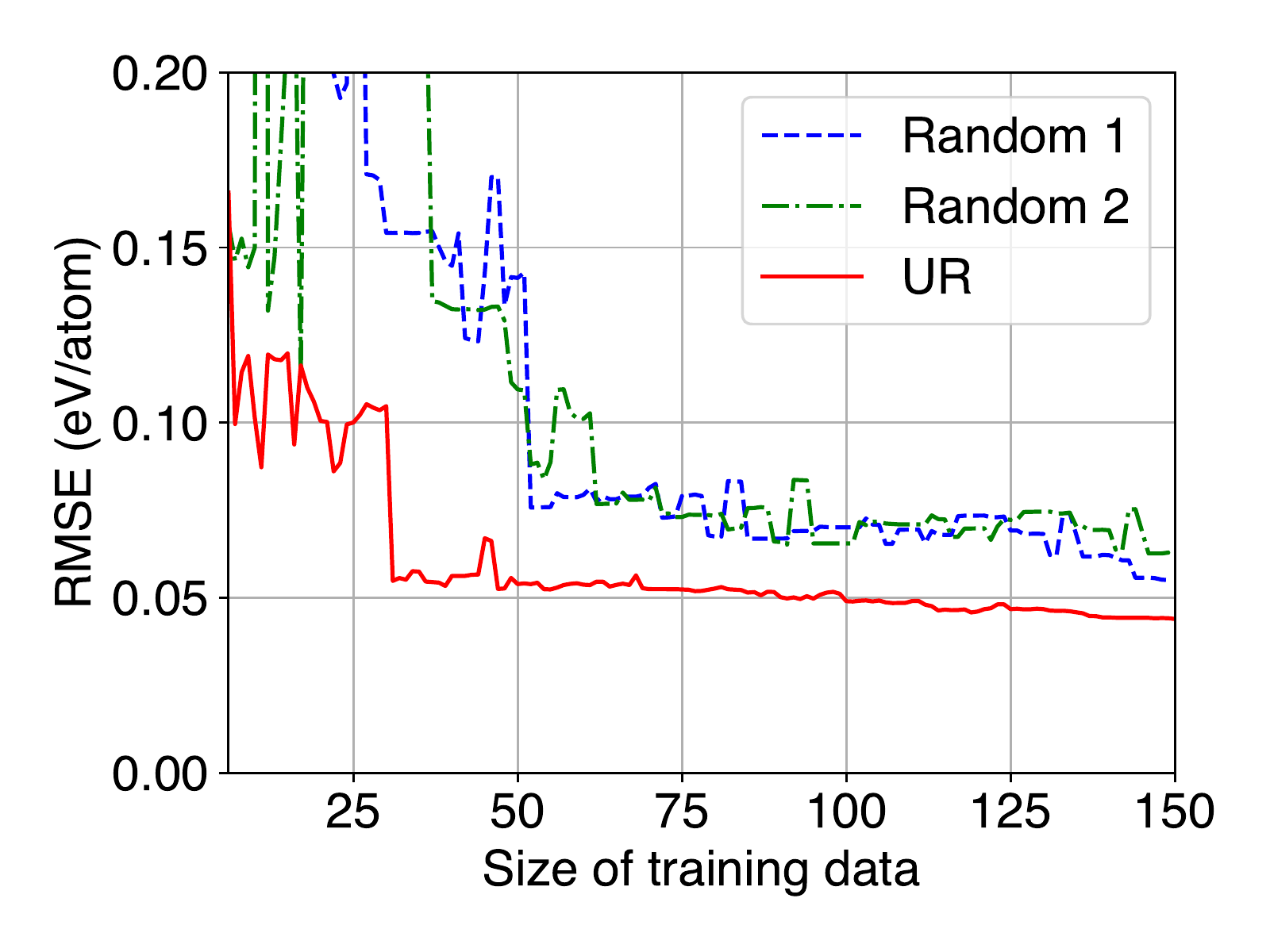}
\includegraphics[scale=0.4]{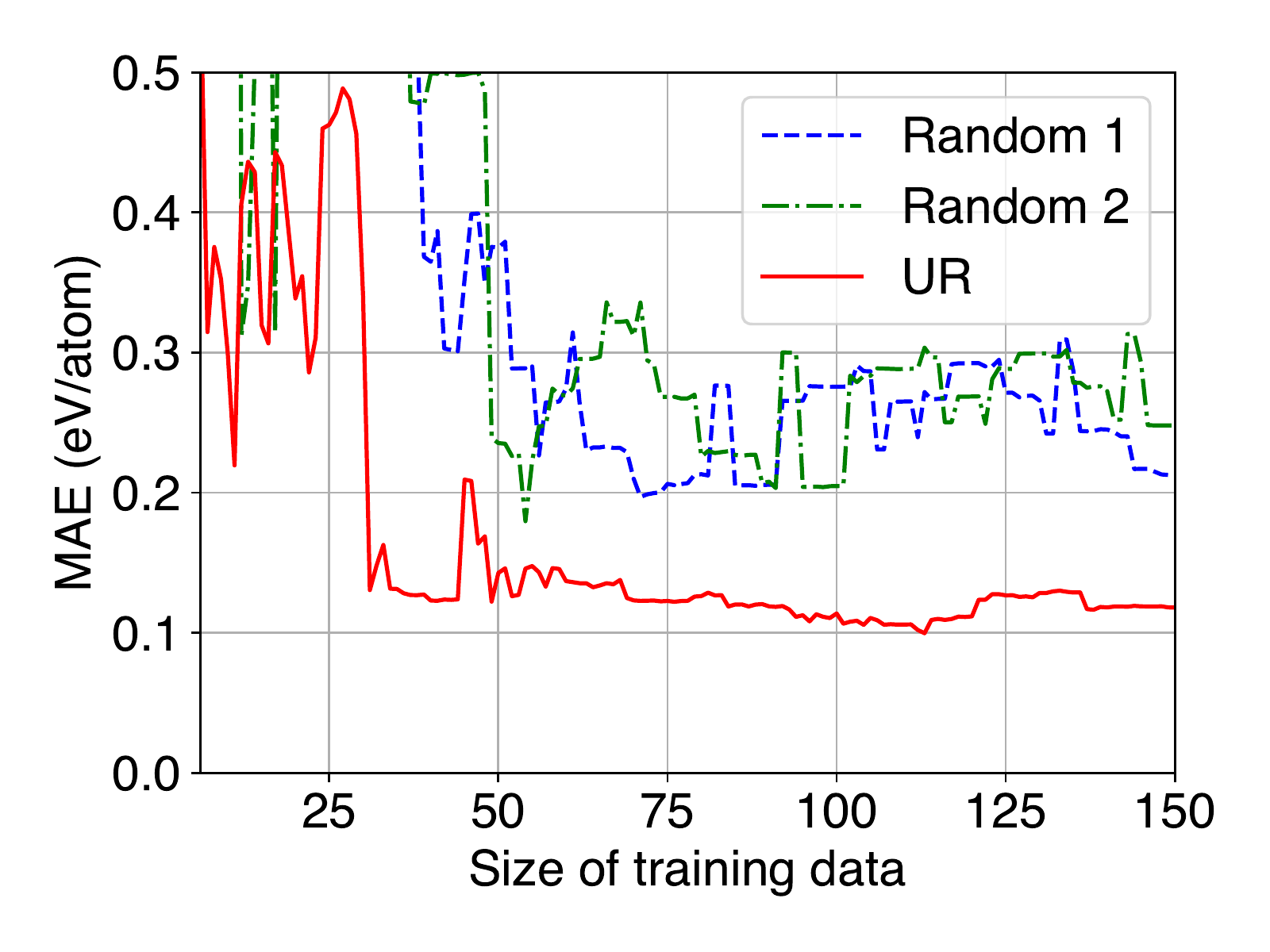}
\caption{
	Transition of prediction error.
	(Left) RMSE (eV/atom) and (Right) MAE (eV/atom).
}
\label{fig:Error}
\end{center}
\end{figure}

The transitions of RMSE and MAE are shown in Fig.~\ref{fig:Error}.
As the dimension of the descriptor is $d = 39$, the errors widely fluctuate 
	when the size of the training data is ~$40$.
It is clear that the UR steadily achieved the lowest errors among the three strategies with respect to RMSE and MAE.

\begin{figure}[t]
\begin{center}
\subfloat[Random~1]{\includegraphics[scale=0.3]{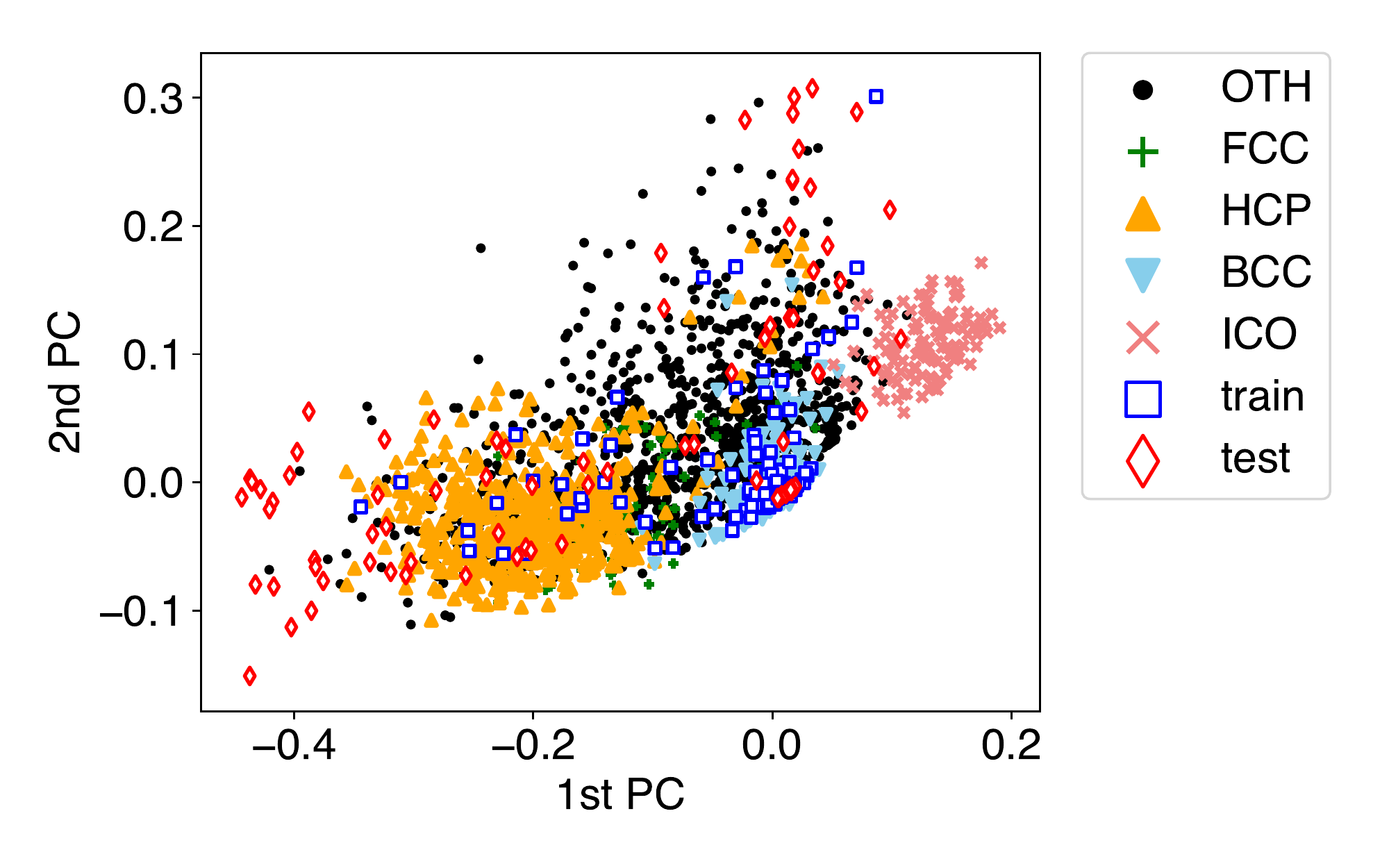}}
\subfloat[UR]{\includegraphics[scale=0.3]{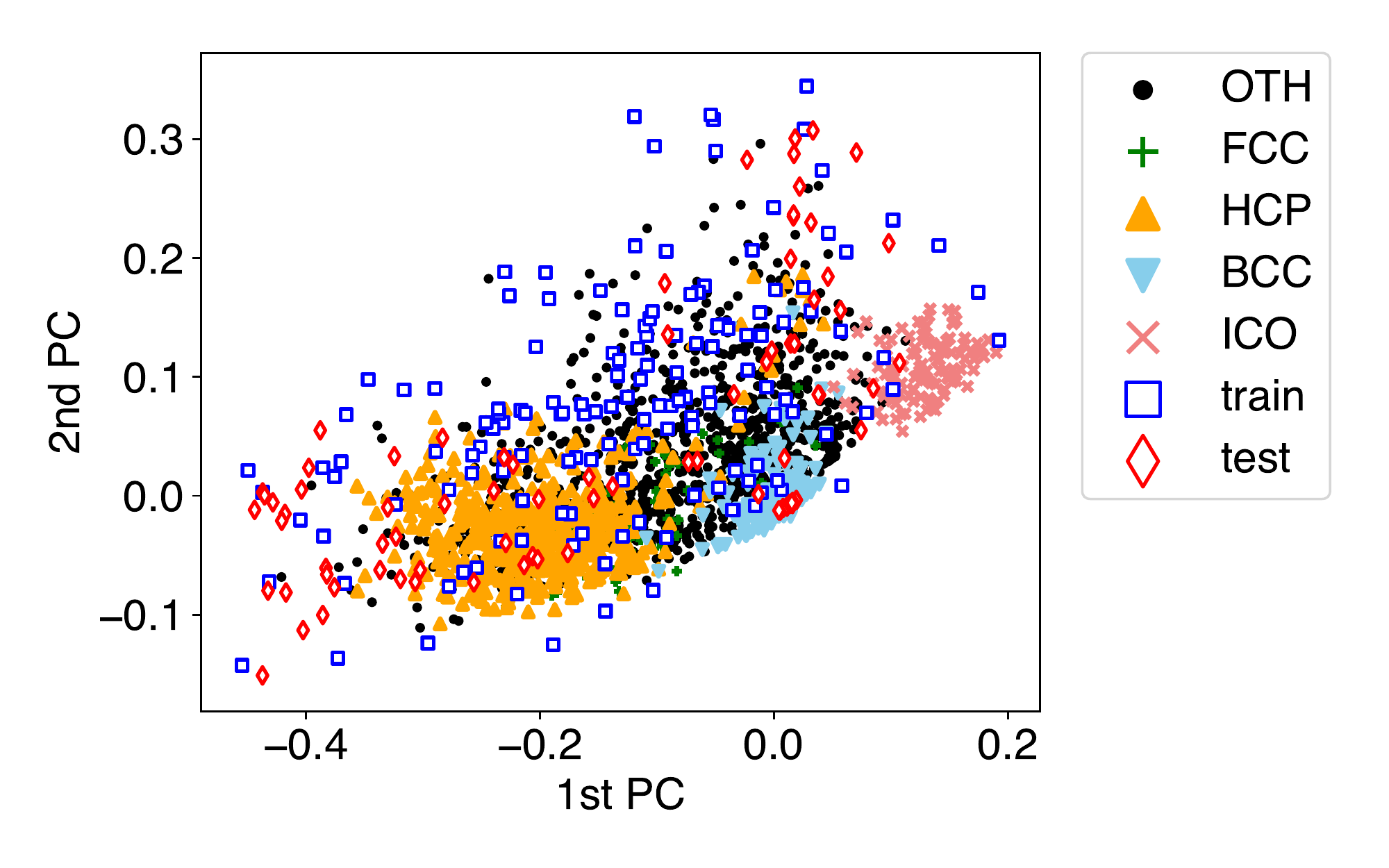}}
\caption{
	Training and test data in the reduced dimensional space created by PCA.
	The data points with the CNA structure type are the same as those in Fig.~\ref{fig:PCA}.
}
\label{fig:PCA_training}
\end{center}
\end{figure}

Figure~\ref{fig:PCA_training} shows the scatter plots of the training and test data in the two-dimensional space created by PCA.
We observe that the test set $\cX_{\rm Test}$ is diversified in the two-dimensional space, and our accuracy analysis 
	covers a variety of structures.
The training instances of UR are widely distributed compared with Random~1, 
	which is concentrated around the center of the plot (Fig.~\ref{fig:PCA_training}~(b)).
As UR attempts to reduce the uncertainty of the sum of all the atoms, it tends to select from a wide range of the input space, 
	which makes the resulting estimation more stable.

\begin{figure}[t]
\begin{center}
\includegraphics[scale=0.4]{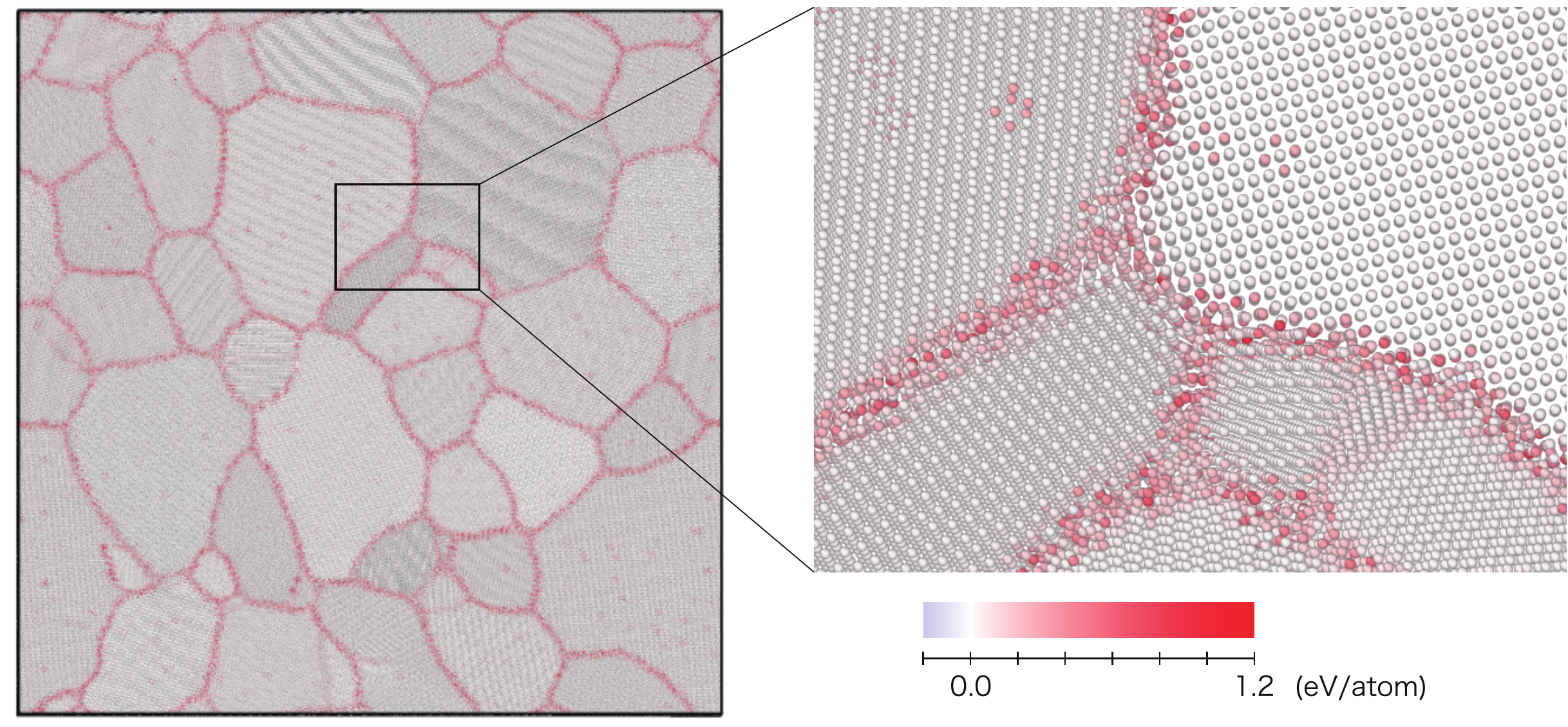}
\caption{Distribution of the predicted atomic energy values. 
	Energy differences from the bulk value are indicated by colors.}
\label{fig:dist_eatom}
\end{center}
\end{figure}

We can plot the distribution of the predicted atomic energy values, as shown in Fig.~\ref{fig:dist_eatom}.
We can observe that the values of atomic energy at the GBs are much larger than those in the bulk region, 
	and the atomic energy of the atoms surrounding the point defects in the bulk region is slightly larger than that of the bulk.

\begin{figure}[t]
\begin{center}
\includegraphics[scale=0.3]{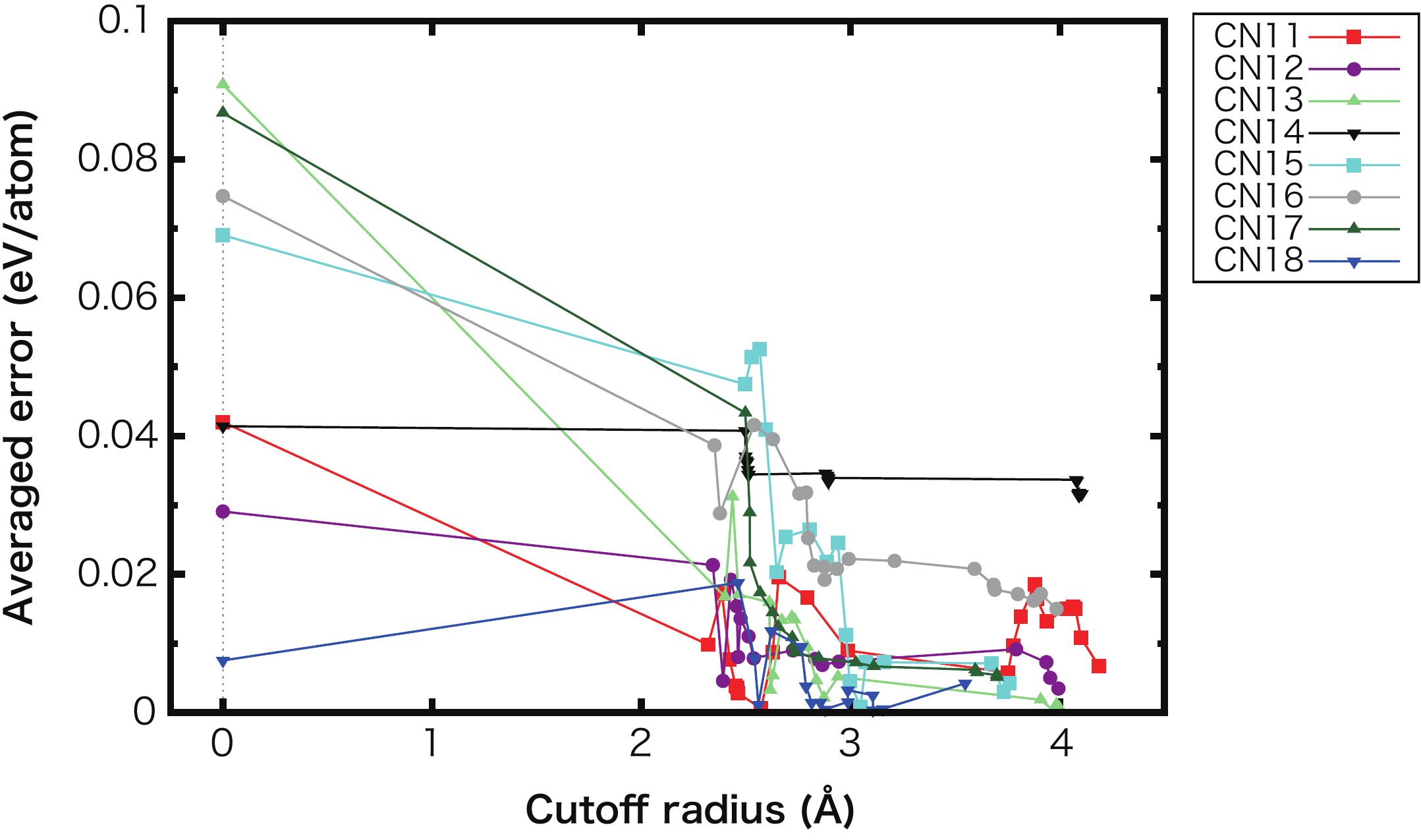}
\caption{Averaged error in the region within the cutoff radius $r_ {c}$ from the central atom.}
\label{fig:dE}
\end{center}
\end{figure}

Various machine-learning-based potentials have been proposed for the high-precision prediction of defect structures.
The transferability of potentials is evaluated by the prediction error, 
	and the total energy of the system is basically used as
	\begin{equation}
		\mathrm{RMSE}=
		\frac{1}{M}\sqrt{\sum_{j=1}^{M} \left |\frac{\Delta E_{j}^{\mathrm{total}}}{N_{j}} \right |^{2}}
		=\frac{1}{M}\sqrt{\sum_{j=1}^{M} \left |\frac{1}{N_{j}}\sum_{i=1}^{N_{j}} \Delta E^{\mathrm{atom}}_{i} \right |^{2}}.
	\end{equation}
Conversely, we evaluate the prediction error using the atomic energy values described in Eq. (\ref{eqn:RMSE}).
Based on the error evaluation using the total energy, the average error is small 
	if most atoms are close to a bulk-like environment.
Therefore, the average error in the region within the cutoff radius $r_ {c}$ from the central atom was evaluated as
	\begin{equation}
		\Delta \tilde{E} (r_{c})=
		\left | \frac{1}{N_{r_{i}<r_{c}}}\sum_{r_{i}<r_{c}}\Delta E^{\mathrm{atom}}_{i} \right |.
	\end{equation}
From the test data set with the coordination number 11$\sim$18, 
	the one with large error was selected as the central atom.
Figure~\ref{fig:dE} shows the $r_{c}$ dependence of $\Delta \tilde{E} (r_{c})$.
For the bcc structure, 
	the halfway value between the second and third coordination shells   
	is 3.46 \AA\ for $a_{\mathrm{bcc}}$=2.87, as described in Eq. (\ref{eqn:rcut}).
The averaged values of $\Delta \tilde {E} $ are 55.16 meV/atom at $r_ {c}$ = 0.0 \AA\ 
	and 11.48 meV/atom at $r_ {c}$ = 3.50 \AA.
As an atom with the coordination number 14 is chosen from the bulk-like region, 
	almost all the surrounding atoms contain the same error because they have the same local environment 
	and the average error does not decrease, even if the number of atoms increases.
From these analyses, we can conclude that our method can be used to predict local energies near defect structures.

In this study, 
	we aim to improve the DFT-level atomic energy of the GB model created using the EAM potential.
As the correlation information between the atomic local environment and the atomic energy is general, 
	there are at least two significant future directions for this study.
One is the possibility of developing atomic-relaxation calculations using the correlation information 
	between the local environment and energy/force field.
We will be then able to predict the stable atomic configuration using DFT calculations 
	based on using the empirical potential.
The other is the possibility of developing the prediction of the atomic configuration and the energy distribution  
	of various lattice defects, including amorphous structures.

\section{Discussions} \label{sec:discussions}

The machine-learning interatomic potentials (MLIPs), 
	such as the neural network potentials (NNP) \cite{Behler2007,Behler2011,Behler2014},
	the Gaussian approximation potentials (GAP) \cite{Bartok2010,Szlachta2014,Bartok2013, 
	Bartok:2015iw,De:2016ia, Deringer2017},
	and the moment tensor potentials (MTP) \cite{Shapeev2016}
	have recently been proposed.
These nonparametric potentials are based on a regression model which is a function of the atomic environments. 
The approximation properties of MLIPs depend on their algebraic form and on the training set
	used to fit them.
For long time-scale MD simulations including rare events,
	the problem of choosing a proper training set is related to the problem of transferability,
	and `the extrapolation problem of interatomic potentials could be solved 
	by reliably predicting on-the-fly whether a potential is extrapolating on a given configuration' 
	(excerpted from Ref. \onlinecite{Podryabinkin2017}). 
The AL approaches, in which DFT calculations for systems containing typically hundreds of atoms are performed 
	only when the potential moves away from known configurations,
	are currently receiving much attention 
	\cite{Frederiksen2004, Csanyi2004, Mueller2012, Artrith2012, Behler2014, Li2015, 
	Botu2015, Bartok2010,Podryabinkin2017, Peterson2017, Bartok2018, Imbalzano2018, Podryabinkin2019, 
	Bernstein2019, Jinnouchi2019a, Jinnouchi2019b, Zhang2019, Vandermause2020, Mueller2020}.
As pointed out in Ref. \onlinecite{Mueller2020},
	the drawback to these approaches is that 
	when large-scale simulations encounter an unfamiliar atomic environment, 
	it is not trivial to model a representative subset of atoms with a relatively expensive method (such as DFT) 
	at reasonable computational cost.
For large-scale systems, DFT calculations cannot be easily performed
	although the local environment for all atoms can be obtained.
Thus, we apply the AL approach to an efficient selection of training data set obtained by DFT calculations
	with small replica cells
	to interpolate an entire system.
The interpolation can be confirmed with PCA analysis as shown in Fig.~\ref{fig:PCA_training}~(b), 
	in which training data (the blue squares) cover the entire structural-descriptor distribution.
The parameters for existing MLIPs are, in practice, fit to the total energy, forces and the stress tensor simultaneously
	for reference structures.
Although the total energy and forces obtained by conventional DFT calculations 
	for small replica cells or cluster models extracted from a large model can be used to fit the parameters,
	this does not necessarily indicate that the atomic-energy prediction is also optimized.
Further, unlike with our method, it is difficult to evaluate empirically the accuracy because the DFT atomic energy is lacking.

The difference in the definitions of AL in our proposed method and on-the-fly MLIP should be noted.
AL has been studied in the statistics and machine-learning community as a strategy 
	to efficiently build an accurate statistical model.
Most approaches are based on evaluating the uncertainty of the estimated model.
A variety of criteria have been proposed to efficiently reduce the uncertainty 
	by ``active selection'' of training data as reviewed in \cite{Settles:2012}.
In contrast, in the materials science community, strategies for adding extra training data 
	during MLIP-based MD simulations are often called on-the-fly AL. 
Unlike the original machine-learning AL, on-the-fly AL cannot select specific training data because 
	atomic configurations are determined through MD simulations.
With respect to our proposed method, we employ the definition of AL from the original machine-learning studies.

In the training data selection in AL, several optimality criteria 
	such as A-optimality, D-optimality, and V-optimality are widely known \cite{Dean:2015}.
UR can be seen as V-optimality, which considers the average or total variance of a pre-specified finite set of $\*x$.
In our problem setting, we are interested only in the atomic-environments $\*x$ that are included in the $\bar{\cX}$ 
	extracted from the GB model, and other atomic-environments need not be taken into consideration.
Therefore, considering the total variance with respect to $\bar{\cX}$ in Eq. \eq{eq:acq}, i.e., V-optimality, 
	is the most direct way to achieve higher accuracy.
For example, A- or D- optimality, which takes into account the uncertainty of the model parameter $\*w$, 
	can be more appropriate when atomic-environments that are not included in the GB model have to be considered.

It must be emphasized that our method is proposed for the training-data selection, 
	and not the algebraic form of interatomic potentials. 
A fundamental assumption for the present AL approach is that 
	the atomic energy is represented by the linear regression of the atomic structural descriptor.
Although we employed the existing SOAP descriptor based on an insight from our prior work \cite{Tamura2017}, 
	any known structural descriptors can be applicable. 
An important advantage of the linear model is that the uncertainty of the estimated model 
	can be evaluated using only the descriptor vector $\*x$ without any DFT result $y$.
For example, the analytical evaluation of exact uncertainty in neural network models can be difficult 
	because of the complicated model definitions, 
	and numerical approaches such as {\it query-by-committee} (e.g., Ref. \onlinecite{Smith:2018}) are often employed.
However, these types of approaches require DFT results to train a model (or a set of models).
Sequential DFT calculations then are often preferred, 
	because the model updated by a larger amount of training data is expected to provide more accurate uncertainty modeling.
Our variance-based criterion for the linear model does not depends on $y$.
 This indicates that DFT results are not necessary for the selection of candidate configurations, 
	and the accuracy of uncertainty evaluation does not depend on the number of DFT results.
Therefore, our UR approach is particularly suitable for massively parallel computers, 
	using which selected configurations can be calculated in parallel.

Another advantage of the linear model is its stability for small training datasets.
We assume that it is difficult to prepare a large amount of training data for large supercells that include structural defects
	since the computational cost of DFT calculations can be a severe restriction.
Although replica supercells, which typically contain, at most, a few hundred atoms, 
	are much easier to calculate when compared with the entire GB model, 
	they can still be expensive to calculate when compared with calculations of simple perfect crystal models.
Therefore, in our problem setting, machine-learning models should provide stable results even 
	with a small training dataset (for example, less than a few hundreds of calculated energy values).
	It is widely known that complicated machine-learning models such as neural networks can cause over-fitting 
	with a small training dataset, but a simpler model is expected to generate more reliable prediction.
This also suggests that the linear model is suitable for predicting the large GB model.

\section{Conclusions} \label{sec:conclusions}

We developed a method that can analyze huge random GB models 
	with the accuracy of DFT calculations using active learning.
It is assumed that the atomic energy is represented 
	by linear regression of the atomic structural descriptor.
Based on the DFT calculations, the atomic energy, called the replica DFT atomic energy, 
	is obtained using a small cell extracted from a huge GB model.
UR in active learning is used to collect efficient training data concerning the atomic energy.
In this method, atomic energy is not needed to search for candidate points; 
	thus, there is no requirement for sequential DFT calculations.
This method is suitable for massively parallel computers 
	that can execute a large number of jobs simultaneously.
We demonstrate the prediction of the atomic energy of a Fe GB model containing one million atoms 
	using the UR approach.
The rate of decrease of the prediction error is further compared with random sampling. 
We conclude that the UR approach with the replica DFT atomic energy is useful for modeling huge GBs
	and will be essential for modeling other structural defects. 

\section*{Data availability}

The datasets generated during the current study and our machine-learning code are available on request.

\section*{Acknowledgments} 

This work was supported by the ``Materials research by Information Integration'' Initiative (MI$^{2}$I) project 
	of the Support Program for Starting Up Innovation Hub of the Japan Science and Technology Agency (JST), 
	MEXT as a social and scientific priority issue (Creation of new functional devices 
	and high-performance materials to support next-generation industries; CDMSI) 
	to be tackled by using post-K computer, MEXT KAKENHI 
	awarded to T.T. (18K04700) and M.K. (17H04694), 
	and PRESTO awarded to M.K. (JPMJPR15N2).
We would like to thank M Kohyama for helpful discussions on the {\it ab-initio} atomic-energy analysis, 
	Y Shibuta for providing data on large-scale MD simulations 
	and R Kobayashi for valuable information on the calculation of the SOAP.

\section*{Author contributions}

T.T. carried out the DFT calculations and M.K. implemented all machine learning methods.
Both authors co-wrote the paper.

\section*{Competing interests}

The authors declare no competing interests.

\section*{References}

\bibliographystyle{unsrt}
\bibliography{ref}

% --- Figures

\clearpage

\clearpage

\clearpage

\clearpage

\clearpage

\clearpage

\clearpage

\clearpage

\end{document}